\documentclass[pre,twocolumn,floatfix]{revtex4}
\usepackage{color}
\usepackage{graphicx}
\usepackage{amsmath}
\usepackage{array}

\newcommand{\m}{\mathcal}

\begin{document}

\title{\bf Onsager approach to 1D solidification problem and its relation to phase field description.}
\author{Efim A. Brener and D. E. Temkin}
\affiliation{Peter Gr{\"u}nberg Institut, Forschungszentrum J{\"u}lich, D-52425 J{\"u}lich, Germany}
\date{\today}
\begin{abstract}
We give  a general phenomenological description of the steady state 1D front propagation problem in two cases:
 the solidification of a pure material and the isothermal solidification of  two component dilute alloys.
  The solidification of a pure material  is controlled by the heat transport in the bulk and  the interface kinetics.
  The isothermal solidification of two component  alloys is controlled by the diffusion in the bulk and the interface kinetics.
 We find that the condition of positive-definiteness of the symmetric Onsager matrix of interface kinetic coefficients still allows an arbitrary sign of the slope of the velocity-concentration line near the solidus  in the alloy problem or of the  velocity-temperature line 
in the case of solidification of a pure material. This result offers a very simple and elegant way 
to describe the interesting phenomenon of a possible  non-single-value behavior of velocity versus concentration which has previously been discussed  by different approaches. We also discuss the relation of this Onsager approach to the thin interface limit of the phase field description.
\end{abstract}
\maketitle
{ \it Introduction.}
In the recent years the phase field approach to solidification problems has attracted the attention of many researches (see, for example, \cite {PE} and references therein). 
It was originally introduced as a mathematical tool to solve the free boundary problem without directly tracking  the interface position. 
However, more recently  it has  also been considered  as a physical model which can bring additional information compared to the sharp interface approach. In particular, it was observed that the general believe, that  steady state 1D front propagation with positive velocity 
\begin{equation}
V=V_0(\Delta_T-1)\,
\end{equation}
is possible only if $\ (\Delta_T-1)>0$ (see, for example \cite{Saito}), is not the general situation. Here $V$, is the steady state front velocity, $V_0$ is the characteristic velocity  which is proportional to the  kinetic growth coefficient;  $\Delta_T=(T_M-T_0)c_p/L$ is the dimensionless undercooling,   
$T_M$ is the melting temperature and $T_0$ is the temperature in the original phase far away from the interface; $c_p$ is the heat capacity 
which is assumed to be the same in both phases; $L$ is the latent heat. 
Karma and Rappel (KR) \cite {KR} introduced the thin interface limit of the phase field description and found that 
  \begin{equation}
\label{linear}
V=\frac{V_0(\Delta_T-1)}{1-a\frac{W V_0}{D_T}}\,
\end{equation}
where $D_T$ is the thermo-diffusion coefficient, $a$ is a positive number of order unity which depends on the details of the model, and $W$
is the thickness of the interface in the phase field description. In the phased field model discussed in  {\cite{KR} there is no any  restriction on the parameter $V_0W/D_T$ and  the velocity may be positive for $(\Delta_T-1) <0$ .  The same result was obtained for the isothermal solidification of alloys by many authors starting from a paper by 
L\"owen et al. \cite{L} in the framework of phase field description and also by Aziz and Boettinger \cite{AB} who use a more phenomenological approach. In the case of alloys the deviation from equilibrium is defined by $\Delta_C=(C_L-C_0)/(C_L-C_S)$ instead of $\Delta_T$. 
In the two phase region of the phase diagram $0<\Delta_C<1$. Here $C_L$ and $C_S$ are the equilibrium concentrations of the initial and growing phase, respectively, and $C_0$ is the concentration of the initial phase far from the interface. 
They found that the steady state growth is possible also inside of the two-phase region of the equilibrium phase diagram.

From the numerous papers on the derivation of the sharp and thin interface limits from a phase field model we should also mentioned 
the work by Elder et al. \cite{E}, Umantsev  \cite{U} and also the paper by  Korzhenevskii, Bausch and Schmitz \cite{KBS} which contain many details and technical points. The basic results of all these descriptions have the structure of Eq. (\ref{linear}) in the vicinity of 
$(\Delta_{T(C)}-1)\ll1$ and eventually lead to the non-single-value behavior of the velocity as a function of the driving force in the case of a negative "kinetic coefficient", Fig.1. In this case the branch which is described by Eq. (\ref{linear})  (dotted line in Fig.1) is linearly unstable (see, for example, \cite {KR}, \cite{L}, \cite{KBS}) while the "high velocity" branch of the mentioned  non-single-value behavior is linearly stable. 
\begin{figure}
\begin{center}
\includegraphics[angle=0,width=0.9\linewidth]{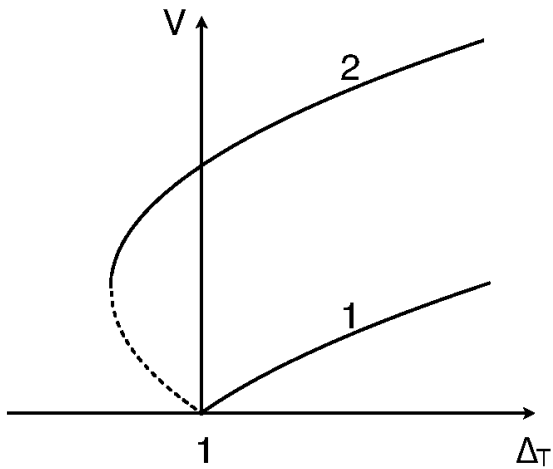}
\caption{Schematic dependence of the steady state velocity $V$ vs. the dimensionless undercooling $\Delta_T$. 
The curve 1 corresponds to the case $aW V_0/D_T <1$ while the curve 2 corresponds to the opposite case, $aW V_0/D_T >1$.}
\label{GeometrySD}
\end{center}
\end{figure}

Qualitatively the same results have been obtained by the numerical solution of 1D motion of the atomically rough interface in binary alloys 
\cite {BT}. In this model instead of the phase field order parameter the authors used the fraction of the atomic places which belongs to the growing phase. This fraction changes from 0 to 1 during the growth. The evolution equations  for this quantity together with the concentration fields in the two phases are given by Eqs. (5.1)-(5.3) in \cite{T81}. The numerical analysis of \cite{BT} has shown that both types of curves in Fig.1 are possible. However, the unstable (dotted line) branch was not seen in this dynamical simulations. 

The  purpose of this Communication  is to give  a general phenomenological description of the steady state 1D front propagation problem in two cases:
i) the solidification of a pure material which is controlled by the heat transport in the bulk and  the interface kinetics;  ii) the isothermal solidification of two component dilute alloys which is controlled by the diffusion in the bulk and  the interface kinetics.  
Describing the interface boundary conditions we use only the general phenomenology of linear non-equilibrium thermodynamics 
in the spirit of the Onsager matrix of kinetic coefficients which has the  proper symmetry and is positive-definite  as required by the second law of thermodynamics. This approach does not assume any specific model of the interface and makes no assumption on its thickness. The only  
requirement, as for any macroscopic theory,  is that   the thickness is small compared  
the macroscopic lengths. We will see that two mentioned problems are very close to each other and can be formally mapped onto each other.
The mentioned restrictions on the Onsager matrix of kinetic coefficients  are not sufficient to determine the sign of the slope of the velocity-concentration line near the solidus  in the alloy problem (or of the velocity-temperature line in the case of solidification of a pure material). 
This result offers a simple way to describe the mentioned above phenomenon of a  non-single-value behavior, Fig.1.  

 The sharp  ($W\rightarrow 0$) and the thin interface limits of the phase field description should lead to the effective macroscopic description with the boundary conditions in the spirit of Onsager relations, where the elements of the Onsager matrix  are expressed in terms of the phase field model parameters. Indeed, these limits really correspond to such a description. However, the mentioned condition of positive-definiteness  of the matrix of kinetic coefficients turns out to be a nontrivial issue for the thin interface limit and  will be discussed in more details.  

{  \it Growth of a pure material with  heat transport.}
We assume that  phase 1 grows at the expense of  phase 2 by a 1D front propagation with the steady state velocity $V$. In the bulk we have the thermal-conductivity equation. In order to write down the boundary conditions at the interface we follow the description and notations given in \cite {BAP}.
\begin{eqnarray}
\label{mu1}
(\mu_2-\mu_1)/T_M&=& \m AV+\m BJ_E \ ,\\
\label{mu2}
(T_2-T_1)/T_M^2&=& \m BV+\m CJ_E \ ,
\end{eqnarray} 
 where $\mu_i$ is the chemical potential of the corresponding phase $i$ at the interface.
  According to the energy conservation at the interface, flux $J_E$ is defined by Eqs. (51)-(52) in \cite{BAP}.
\begin{eqnarray}
\label{G1}
-\lambda_1\nabla T_1&=&VT_MS_1-J_E ,\\
\label{G2}
-\lambda_2\nabla T_2&=& VT_MS_2-J_E \ ,
\end{eqnarray} 
Here $S_1(T_1)$ and $S_2(T_2)$ are the entropies of two phases and $\lambda_i$ is the thermoconductivity of the phase $i$. The elements of the Onsager matrix, which is symmetric and positive-definite, obey the conditions $\m A,\m C>0$ and $\m B^2<\m A\m C$.  $R_K=\m CT_M^2$ is the  Kapitza resistance and the cross coefficient $\m B$ describes the way the two entropies are shared between the two sides of the interface during growth (for a more detailed discussion of the physical meaning of the different Onsager coefficients in this case see \cite {BAP}).

For the steady state one-dimensional problem $\nabla T_1=0$ and $T_1=T_0+L/c_p$ where $L=T_M[S_2(T_M)-S_1(T_M)]$ is the latent heat 
and $c_p$ is the heat capacity; $T_0$ is the temperature in the original phase far away from the interface. We note that in order to obtain the relation $T_1=T_0+L/c_p$ one should expand the entropies near the equilibrium temperature $T_M$ in the energy conservation condition $(\lambda_1\nabla T_1-\lambda_2\nabla T_2)=VT_M[S_2(T_2)-S_1(T_1)]$. Now expanding the difference of the chemical potentials near the equilibrium temperature $T_M$, we find
\begin{equation}
\mu_2(T_2)-\mu_1(T_1)=  (S_2-S_1)(T_M-T_1) +S_2(T_1-T_2)\ ,
\end{equation}
 and finally we get 
 \begin{equation}
 \label{linear7}
V=\frac{L^2(\Delta_T-1)}{c_pT_M^2[\m A+\m CT_M^2S_1S_2+\m BT_M(S_1+S_2)]}\ ,
\end{equation}
where $\Delta_T=(T_M-T_0)c_p/L$. We have used the  usual notation for solidification of pure materials.  We see that the sign of $(\Delta_T-1)$
in general is not determined by the Onsager restriction $\m B^2<\m A\m C$. However, it is well defined in two cases: i)$\m B=0$ and ii) in the  "isothermal" case, $T_1=T_2$. In the later case  the growth rate is controlled by the  "isothermal" kinetic coefficient which is strictly positive due to the mentioned restriction, $\m B^2<\m A\m C$ \cite{BAP}:

\begin{equation}
 \label{linear8}
V= \frac{(\mu_2-\mu_1)}{T_M\m A[(1-\m B^2/(\m A\m C)]}=\frac{L^2(\Delta_T-1)}{c_pT_M^2\m A[(1-\m B^2/(\m A\m C)]}\ ,
\end{equation}

Karma and Rappel obtained in their thin interface limit $\m B=\m C=0$ and a coefficient $\m A$ which may  even be negative
($\beta$ in their notation). They  discussed this "counterintuitive" issue and gave some natural explanation for this phenomenon. We will return to this point later.

{  \it Isothermal alloy solidification in the dilute limit.}
We discuss the steady state propagation of a 1D front  with  velocity $V$ during solidification of  a two component alloy at a given temperature $T$. The concentration of B atoms is $C_1(x)$ in   phase 1 and $C_2(x)$ in  phase 2. In the bulk these concentrations are described by diffusion equations with diffusion coefficients $D_1$ and $D_2$. 
In order to write down the boundary conditions in this case we use the same phenomenological approach but adapted to the alloy situation. Onsager relations connect two fluxes $J_A$ and $J_B$ (at the boundary) of atoms  $A$ and $B$ to two driving forces $\delta\mu_A$ and $\delta\mu_B$ which are usual differences in chemical potentials at the boundary. 
 While  the bulk is described  by diffusional equations for the concentration fields  for each phase, 
 we still need three boundary conditions at the interface. One is the  conservation of B atoms at the interface. We have  also to relate the two concentrations $C_1$ and $C_2$ on both sides of the interface to the growth velocity and gradients of the concentrations.  In the equilibrium these two concentrations are just the liquidus and solidus concentrations. When the velocity is finite these two concentrations deviate from the equilibrium values. We write (see, for example, \cite{caroli} and references therein):  
\begin{eqnarray}
\label{mu3}
\delta\mu_A/T&=& \m AJ_A+\m BJ_B \ ,\\
\label{mu4}
\delta\mu_B/T&=& \m BJ_A+\m CJ_B \ ,
\end{eqnarray} 
This Onsager matrix should be positive-definite: $\m A$ and $\m C$ must be positive and $\m B^2<\m A\m C$.
According to  the conservation of  $B$ atoms  at the interface we also have \cite{T81}
\begin{eqnarray}
-D_1\nabla C_1&=&VC_1-J_B \ ,\\
-D_2\nabla C_2&=& VC_2-J_B\ , \\
                V&=&J_A+J_B \ .
\end{eqnarray} 
Eq. (14) is written for  substitutional alloys. For interstitial alloys $V=J_A$. 
For dilute alloys the chemical potential are \cite{LL}
\begin{eqnarray}
\delta\mu_A/T&= &(C_1-C_2) +(C_L-C_S) \ ,\\
\delta\mu_B/T&=& \ln(C_2/C_1)+\ln(C_S/C_L) \ ,
\end{eqnarray} 
Here  phase 1 grow at the expense of phase 2. $C_1$ and $C_2$ are the concentrations of $B$ atoms at the interface and $C_S$ and $C_L$ are their equilibrium values; $(C_L-C_S)\sim(T_M-T)/T$ is proportional to the deviation of the temperature from its equilibrium value for a pure $A$ material.  $D_1$ and $D_2$ are the diffusion coefficients. 

According to the mass conservation at the interface for the steady state 1D growth we have $J_A= V(1-C_1)$ and $J_B=VC_1$ because there is no gradient in the growing phase 1. These relations are written for the 
substitutional alloys. For the interstitial alloys $J_A=V$. However, in the dilute limit there is no difference between these two alloys because $C_1\ll1$ and can be neglected in the expression for $J_A$ for the substitutional alloys. 
Moreover, the global mass conservation requires that $C_1=C_{0}$, where $C_{0}$ is the concentration in  original phase 2 far away from the interface. 
Solving the resulting system of equation we find the transcendental  relation between velocity $V$ and the initial concentration $C_0$:
\begin{equation}
\label{ln}
\ln \left[{\frac{C_S }{C_L}\left [1+\frac{C_L-C_S}{C_0}-V(\m A/C_0+\m B)\right ]}\right]= V[\m B+\m C C_0]. 
\end{equation}
If the concentration $C_0$ is close to $C_S$ and the velocity $V$ is small we find, expanding logarithm up to linear order in $(C_0-C_S)$ and $V$, 
\begin{equation}
\label{linear1}
V=\frac{(C_L-C_S)(C_S-C_0)}{C_S[\m A+\m CC_LC_S+\m B(C_L+C_S)]}
\end{equation}
For the general case of not dilute alloys this equation reads
\begin{equation}
\label{linear2}
V=\frac{[f_1''(C_S)/T](C_L-C_S)(C_S-C_0)}{\m A(1-C_L)(1-C_S)+\m CC_LC_S+\m B[(C_L+C_S)-2C_LC_S]}\ ,
\end{equation}
where $f_1''(C)$ is the second derivative of the free energy  $f_1(C)$ of the growing phase 1 with respect to the concentration. 
From this expression it is clear that in the presence of the cross coefficient $\m B$ the sign of $(C_S-C_0)$ is not determined by the  
condition $\m B^2<\m A\m C$ and also depends on $C_L$ and $C_S$. Moreover, if the sign in the square brackets of Eq. (\ref {linear1}) is negative and $C_0>C_S$ for 
small positive velocity $V$ then we  find for $C_0=C_S$ apart from  the  solution $V=0$  second solution with positive $V$. If the expression in the square brackets is negative but small, we can expand
the  logarithm up to linear order in $(C_0-C_S)$ and up to quadratic order in $V$ and find:
\begin{eqnarray}
\label{nonlinear1}
\frac{(C_L-C_S)(C_S-C_0)}{C_S} &=& 
V[\m A+\m CC_LC_S+\m B(C_L+C_S)] \nonumber \\
&&+V^2[\m A+\m BC_S]^2/(2C_L)
\end{eqnarray}
This expression shows that with increasing $V$ the curve $V=V(C_0)$ first goes into the two-phase region, then turns back having another 
solution with finite velocity at $C_0=C_S$ and then goes into the one phase region (see Fig.1). Eventually, for $C_0\rightarrow 0$ the velocity, according to Eq. (\ref{ln}), becomes  
$V=(C_L-C_S)/\m A\sim (T_M-T)/(T\m A)$ as for the solidification of a pure  material. 

First of all,  we would like to mention the clear analogy between two discussed problems. From the basic equations we see this analogy  if we relate 
$V\rightarrow J_A$, $J_E\rightarrow J_B$, $T_MS_{1(2)}\rightarrow C_{1(2)}$, $\delta\mu\rightarrow\delta\mu_A$, $\delta T/T_M^2\rightarrow\delta\mu_B/T$ and apart from some thermodynamical prefactors $(\Delta_T-1)\rightarrow (C_S-C_0)$. 
Then, the case $\delta T=0$  in the pure material problem corresponds to zero values of $\delta\mu_B$. This, in turn, corresponds to a frequently used assumption that the partition coefficient $k=C_1/C_2$ equals to its equilibrium value $k_0=C_S/C_L$.  In this case as in the pure material problem  stationary growth is possible only in the one phase region of the phase diagram. Actually it seems that this result is in agreement with the phenomenological description of \cite{AB} and also \cite{KBS}.

{ \it  Discussion and conclusion: Thin interface limit of phase field models vs. Onsager approach.}
We discuss the thin interface limit using the KR description for the temperature field for a flat interface. The corresponding problem for alloys leads to basically the same results (see, for example, \cite{E, KBS,K}). Originally  it was designed  to increase computational power of the method by using larger values of the interface width $W$ and to mimic local equilibrium boundary conditions \cite{KR}. Let us have a closer look at this limit from more physical prospectives.
 In the thin interface limit of \cite{KR} the temperature distribution $T(x)$ close to the interface is given by $T_i(x)= T(0)+G_ix$ where $G_i$ is the  temperature gradient in the i-th phase ($i=1,2$) at $x=0$. At $x=0$ the temperature $T_1=T_2=T(0)$ and in this description the Kapitza jump is absent, $T_1-T_2=0$.   One should note that the value of $T(0)$ in this linear extrapolation procedure is  different from the real value of the smooth temperature field at the middle of the interface obtained by the phase field simulations. The given linear extrapolation of the temperature field reasonably coincides with direct phase field results only for $H\gg |x|\gg W$ where $W$ is the width of the phase field and $H\gg W$ is some macroscopic length scale.  KR  derived a kinetic boundary condition which relates the effective temperature $T(0)$ and the growth velocity $V$ by the  kinetic coefficient $\m A_{KR}$: $[T_M-T(0)]L/T_M^2=\m A_{KR}V$. Using the asymptotic matching procedure they obtained  that the kinetic coefficient has the following structure: 
\begin{equation}
\label{kin}
\m A_{KR}=\frac{L^2}{T_M^2c_p} \left(\beta_0-a\frac{W}{D_T}\right),
\end{equation}
where $\beta_0=1/V_0>0$ is the KR kinetic coefficient in the sharp interface limit ($W\rightarrow 0$) and $a$ is a positive numerical factor of the order of unity which depends on some tiny details of the specific phase field model.  The second negative term is due to the finite thickness $W$ of the interface and the described linear extrapolation procedure. We also note that in this description the other Onsager coefficients vanish, $\m B=\m C=0$ in both sharp and and thin interface limits. KR checked that for the steady state 1D growth, the analytical prediction, Eq. (\ref{linear7}) with the obtained value of $\m A_{KR}$ and $\m B=\m C=0$, is in good agreement with direct numerical simulations of the phase field model. However, there is a subtle physical point concerning the interpretation of  $\m A$, which may become negative with some choice of phase field model parameters. As correctly mentioned by KR, this conclusion may appear at first sight thermodynamically inconsistent. However, as it has been already mentioned, the temperature $T(0)$ is not a real temperature inside of the interface and deviates strongly from the temperature obtained by phase field simulation, which is below $T(0)$. 

Let us discuss this nontrivial point in more details. We can imagine an extended interface with the thickness $2\delta$ with two boundaries located at $x=\pm\delta$. We  emphasize that this length scale $\delta$  is different from the phase field interface width $W$ and  for the moment arbitrary, still being much smaller than any relevant macroscopic length scales. We can easily derive the corresponding matrix of Onsager coefficients 
using the values of $T$ and $\mu$ at the two boundaries of the extended interface 
as $T_1=T(0)-G_1\delta$ and $T_2=T(0)+G_2\delta$, and $\mu_1(T_1)$ and $\mu_2(T_2)$. Using Eqs. (\ref{G1})-(\ref{G2}) we express the temperature gradients $G_i$ in terms of $J_E$ and $V$, and using Eqs. (\ref{mu1})-(\ref{mu2}) we finally find the renormalized values of the Onsager coefficients 
\begin{eqnarray}
\label{renorm1}
\m A(\delta)=\m A_{KR}+\m C(\delta)T_M^2(S_1^2+S_2^2)/2, \\
\m B(\delta)=-\m C(\delta)T_M(S_1+S_2)/2,\\
\label{renorm3}
\m C(\delta)T_M^2=2\delta/\lambda,
\end{eqnarray}
where,  we have assumed that $\lambda_1=\lambda_2=\lambda$ as in \cite{KR}. Few remarks are in order.

 i)  The steady state result, Eq. (\ref{linear7}), is invariant with respect to this renormalization of the Onsager coefficients, i.e. independent of $\delta$. 
It means that  this $\delta$-family of Onsager matrixes  is in good agreement with numerical simulations of the phase field model as well as the original KR case, $\delta=0$.

ii) With the choice $\delta>2aW$ the matrix of Onsager coefficients becomes positive-definite, $\m A,\m C>0$ and $\m A\m C>\m B^2$, for arbitrary parameters of the phase field model. This result has  a clear physical meaning. For $\delta\gg W$ we discuss only the range of $|x|$ where the used linear extrapolation of the temperature field is in agreement with the temperature field obtained by the phase field, while for $\delta\ll W$ the temperature 
at the boundaries strongly deviates from the phase field description, which is fully thermodynamically consistent by itself. In other words, for 
$\delta\gg W$ the obtained matrix of kinetic coefficients does describe real physical dissipation in the region $\delta$, while for $\delta\ll W$ 
this "effective" matrix does not describe the real physical dissipation, but still leads to the correct expression for the steady state growth velocity.

This possible renormalization with $\delta$, much smaller than any macroscopic length scale $H$,  is not specific only to the phase field models and represents a small effect of the order of $\delta/H\ll1$. It has the same structure as the "negative" phase field effects $W/H$. 
The   ideology of any macroscopic description relies  on 
this small parameter as an expansion parameter of the theory. 
These corrections should be irrelevant in the general case of the diffusional transformation where the bulk dissipation plays the major role (for example, in the case of dendritic growth at small undercooling). 
We have seen, however, that  in the specific problem of steady state 1D front propagation  this 
small term (proportional to $W$) is responsible for the sign of the slope in the phase field model description. This happens because the bulk dissipation (being still much larger than the interfacial dissipation) just bring us to the vicinity of point $\Delta_T=1$ and does not contribute to the  slope. In this case the growth velocity is entirely controlled by the interface kinetics. We note that the interpretation of  the nontrivial behavior in the  vicinity of $\Delta=1$  due to sufficiently negative values of the phenomenological cross coefficient $\m B$ does not assume any specific model of the interface. At the same time, the explanation suggested by the phase field modeling explicitly takes  inhomogeneities of the temperature and concentration fields, on the scale of finite interface thickness, into account. 

In other words, there is no doubt about thermodynamic consistency of the phase field model for arbitrary values of the parameter $V_0W/D_T$ apart from the obvious restrictions, $V_0>0$ and $D_T>0$. However, the interpretation of the thin interface limit and its relation to the matrix of dissipative Onsager coefficients should be taken with care.  We illuminate this warning by the following additional example. Let us assume that initially the two-phase system is at some temperature $T$  slightly below the melting temperature $T_M$. This system  evolves towards equilibrium with a solidification velocity $V$  that decays as   $ V\sim t^{-1/2}$ at large time $t$. This behavior would be observed in direct phase field simulations for arbitrary parameters of the model independent of  the sign of effective kinetic coefficient, Eq. (\ref{kin}). A slightly different but close in spirit non-stationary  evolution has been discussed in \cite{KR} confirming this behavior. However, if one solved this problem not by a direct phase field simulation but by solving the free boundary problem with effective boundary conditions described by the the matrix of kinetic coefficients, $\m A=\m A_{KR}$ and B=C=0 (the thin interface limit of \cite{KR}), the result would be very different if $\m A_{KR}<0$. The system would melt, instead of of being solidified, exhibiting strong instabilities and would never  reach the described physical attractor. 
On the other hand, if one solved the same problem using the renormalized positive-definite matrix of Onsager coefficients, Eqs. (\ref{renorm1}-\ref{renorm3}), the result would be basically the same as in direct phase field simulations and physically relevant. Therefore, we conclude that the interpretation of the thin interface limit of \cite{KR} as  the correspondence between the phase field description and the classical macroscopic approach is incorrect for the wide class of non-stationary problems if  $\m A_{KR}<0$. However, the renormalized positive-definite matrix of Onsager coefficients leads to such a correspondence in the macroscopic limit for arbitrary $\m A_{KR}$. 

  Finally, we would like to address one more point. The phase field model of \cite{KR}  contains less independent  parameters to describe the kinetic properties of the interface (only $\m A$ or $\beta_0$) than  is allowed by the general phenomenology ($\m A, \m B, \m C$).  While an independent parameter $\m C$ can be  introduced in a slightly modified version of the phase field model, the introduction of the independent cross coefficient $\m B$ is a serious problem.  
As pointed out in \cite{Sekerka98}, according to Curie's principle \cite{GM62},  there can be no kinetic coupling between the scalar non-conserved phase field order parameter $\phi$ and vectorial diffusional fluxes of the conserved quantities  energy and/or concentration. Thus, 
one should not expect an independent cross coefficient $\m B$ to appear in the effective boundary conditions, Eqs. (\ref {mu1},\ref {mu2}) and Eqs. (\ref {mu3},\ref{mu4}) . However, in the general case of the phenomenological macroscopic description, we do not doubt the existence  of such a kinetic coupling at the interface between the normal growth velocity and normal diffusional fluxes through the interface. It is  conceivable   that this coupling can be introduced in  modified versions of the phase field model where ${\bf\nabla}\phi/|{\bf\nabla}\phi|$,   the unit vector normal to the interface, can be used  to produce the corresponding vectorial quantities.  This issue may also be relevant to the  anti-trapping current introduced in some non-variational versions of the phase field model \cite{K,KP} for different purposes. The anti-trapping current introduces  a new kinetic coefficient and uses the unit vector normal to the interface.  To use this idea for the description of the cross effect of the interface kinetics in  phase field models, one should carefully consider  the necessary symmetry which is obligatory for this cross effect.  A more detailed discussion of this question is far beyond the scope of this paper. 

{\it Acknowledgment}. 
We acknowledge the support of the Deutsche Forschungs- gemeinschaft under Project  SFB 917.

\end{document}